\pdfoutput=1
\documentclass[]{spie}  

\newcommand{\ukrts}{$\mu$K$\sqrt{\textrm{s}}$}

\usepackage{amsmath,amsfonts,amssymb}
\usepackage{graphicx}
\usepackage[colorlinks=true, allcolors=blue]{hyperref}
\usepackage{siunitx}
\usepackage{booktabs}
\usepackage{color}
\usepackage{tablefootnote}

\title{BICEP3 focal plane design and detector performance}

\author[a]{H. Hui}
\author[b]{P. A. R. Ade}
\author[c,d]{Z. Ahmed}
\author[e]{K. D. Alexander}
\author[f]{M. Amiri}
\author[e]{D. Barkats}
\author[g,h]{S. J. Benton}
\author[e]{C. A. Bischoff}
\author[a,i]{J. J. Bock}
\author[e]{H. Boenish}
\author[e]{R. Bowens-Rubin}
\author[e]{I. Buder}
\author[j]{E. Bullock}
\author[e,k]{V. Buza}
\author[e]{J. Connors}
\author[a,l,m]{J. P. Filippini}
\author[n]{S. Fliescher}
\author[d,c]{J. A. Grayson}
\author[f]{M. Halpern}
\author[e]{S. Harrison}
\author[o]{G. C. Hilton}
\author[a]{V. V. Hristov}
\author[d,c,o]{K. D. Irwin}
\author[d,c]{J. Kang}
\author[e]{K. S. Karkare}
\author[d,c]{E. Karpel}
\author[a]{S. Kefeli}
\author[d,c]{S. A. Kernasovskiy}
\author[e,k]{J. M. Kovac}
\author[d,c]{C. L. Kuo}
\author[p]{E. M. Leitch}
\author[a]{M. Lueker}
\author[i]{K. G. Megerian}
\author[d,c]{V. Monticue}
\author[d,c]{T. Namikawa}
\author[g,q]{C. B. Netterfield}
\author[i]{H. T. Nguyen}
\author[a,i]{R. O'ÂÂBrient}
\author[d,c]{R. W. Ogburn IV}
\author[n,j]{C. Pryke}
\author[o]{C. D. Reintsema}
\author[e]{S. Richter}
\author[n]{R. Schwarz}
\author[e]{C. Sorensen}
\author[n,p]{C. D. Sheehy}
\author[a,i]{Z. K. Staniszewski}
\author[a]{B. Steinbach}
\author[a,r]{G. P. Teply}
\author[d,c]{K. L. Thompson}
\author[d,c]{J. E. Tolan}
\author[b]{C. Tucker}
\author[i]{A. D. Turner}
\author[e,s,p]{A. G. Vieregg}
\author[d,c]{A. Wandui}
\author[i]{A. C. Weber}
\author[f]{D. V. Wiebe}
\author[n]{J. Willmert}
\author[t,d,c]{W. L. K. Wu}
\author[d,c]{K. W. Yoon}

\affil[a]{Department of Physics, California Institute of Technology, Pasadena, CA 91125, USA}
\affil[b]{School of Physics and Astronomy, Cardiff University, Cardiff, CF24~3AA, United Kingdom}
\affil[c]{Kavli Institute for Particle Astrophysics and Cosmology, SLAC National Accelerator Laboratory, Menlo Park, CA 94025, USA}
\affil[d]{Department of Physics, Stanford University, Stanford, CA 94305, USA}
\affil[e]{Harvard-Smithsonian Center for Astrophysics, Cambridge, MA 02138, USA}
\affil[f]{Department of Physics and Astronomy, University of British Columbia, Vancouver, BC, V6T~1Z1, Canada}
\affil[g]{Department of Physics, University of Toronto, Toronto, ON, M5S 1A7, Canada}
\affil[h]{Department of Physics, Princeton University, Princeton, NJ 08544, USA}
\affil[i]{Jet Propulsion Laboratory, Pasadena, CA 91109, USA}
\affil[j]{Minnesota Institute for Astrophysics, University of Minnesota, Minneapolis, MN 55455, USA}
\affil[k]{Department of Physics, Harvard University, Cambridge, MA 02138, USA}
\affil[l]{Department of Physics, University of Illinois Urbana-Champaign, Urbana, IL 61801, USA}
\affil[m]{Department of Astronomy, University of Illinois Urbana-Champaign, Urbana, IL 61801, USA}
\affil[n]{School of Physics and Astronomy, University of Minnesota, Minneapolis, MN 55455, USA}
\affil[o]{National Institute of Standards and Technology, Boulder, CO 80305, USA}
\affil[p]{Kavli Institute for Cosmological Physics, University of Chicago, Chicago, IL 60637, USA}
\affil[q]{Canadian Institute for Advanced Research, Toronto, ON, M5G~1Z8, Canada}
\affil[r]{Department of Physics, University of California at San Diego, La Jolla, CA 92093, USA}
\affil[s]{Department of Physics, Enrico Fermi Institute, University of Chicago, Chicago, IL 60637, USA}
\affil[t]{Department of Physics, University of California, Berkeley, CA 94720, USA}

\authorinfo{Further author information: (Send correspondence to H. Hui)\\H. Hui: E-mail: hhui@caltech.edu, Telephone: 1 626 395 2023}

\begin{document} 
\maketitle

\begin{abstract}
BICEP3, the latest telescope in the BICEP/Keck program, started science observations in March 2016. It is a 550mm aperture refractive telescope observing the polarization of the cosmic microwave background at \SI{95}{GHz}. We show the focal plane design and detector performance, including spectral response, optical efficiency and preliminary sensitivity of the upgraded BICEP3. We demonstrate \SI{9.72}{\ukrts} noise performance of the BICEP3 receiver.

\end{abstract}


\section{INTRODUCTION}

Measurements of the polarization of the Cosmic Microwave Background provide key information to further our understanding of the early universe. The $\Lambda$CDM model predicts a E-mode polarization pattern in the CMB at the level of a few $\mu$K and arc-minute B-mode polarization arises from gravitational lensing of E-mode power by the large scale structure of the universe. But inflationary gravitational waves may be a source of degree scale B-mode polarization and a detection of such signal can be use to constrain the tensor-scalar ratio $r$ and place limits on the energy scale and potential from inflation~\cite{Kamionkowski15}. However, several galactic mechanisms can generate B-mode foregrounds; to disentangle the cosmic signal from galactic ones, we need to probe the polarization of the CMB at multiple frequencies with high sensitivity.

The BICEP/ Keck team has deployed multiple telescopes to the South Pole since 2006; we use small aperture, refracting telescopes with high sensitivity receivers to map the degree scale B-mode signal. The Keck Array is in its fifth season, currently observing at \SI{150}{GHz} and \SI{220}{GHz} and previously observed at \SI{95}{GHz}, \SI{150}{GHz} and \SI{220}{GHz}~\cite{Ogburn12} optical bands with 5 BICEP2 style cameras. BICEP3 is the latest addition to this program and was first deployed to the South Pole Station in 2015. It is a \SI{550}{mm} aperture, on-axis, refractive polarimeter designed to observe at \SI{95}{GHz}. During the first observing season, the focal plane was only partially filled with 1152 detectors~\cite{Wu15}, whereas in this year's observing season, the instrument is complete with 2560 detectors. Combining data with BICEP2/Keck, Planck and South Pole Telescope, BICEP3 is projected to set upper limits on $r < 0.03$ at $95 \%$ confidence. Ref.~\citenum{Grayson16} shows an overview of the BICEP3 telescope design and observing strategy.

\section{Focal Plane Design}

BICEP3 uses the same antenna-coupled TES bolometer architecture as BICEP2/ Keck, but we adapted a new modular housing design that allows us to fill larger optically illuminated areas and for easier detector repair and replacement. This section shows the designs of both the detectors and the focal plane module.

\subsection{Antenna-coupled transition-edge sensor (TES)}

\begin{figure} [ht]
   \begin{center}
   \begin{tabular}{c}
   \includegraphics[height=6cm]{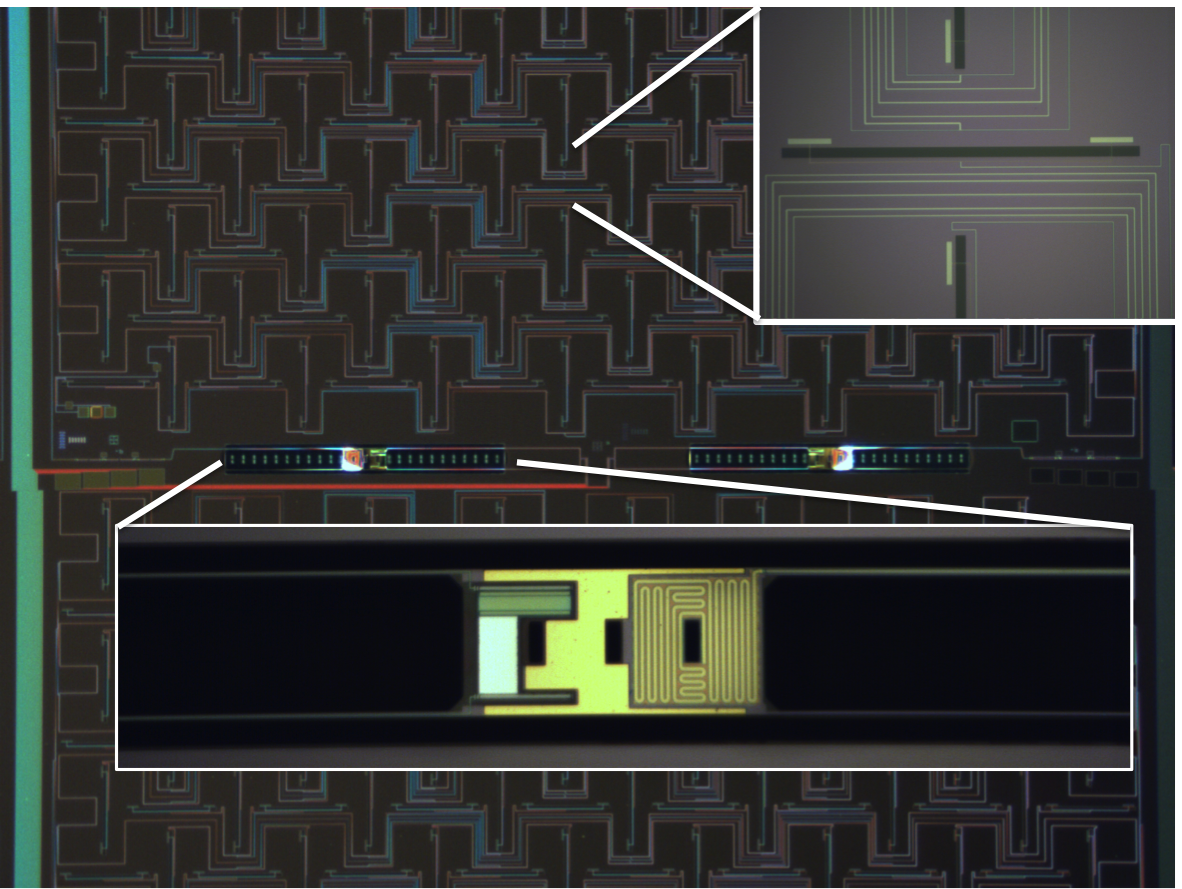}
   \end{tabular}
   \end{center}
   \caption[TES]
   { \label{fig:TES} 
This figure shows the slot antenna array. Each pixel consists of two collocated 8 by 8 orthogonally polarized antenna arrays. Top right inset is a zoom showing the slot antennas for both polarizations. Bottom inset shows the TES bolometer. The Al and Ti TESes are at the left, gold microstrip termination is at right, and a thick layer of gold in the middle ensures thermal stability.}
   \end{figure}

We couple optical power to our detectors through pairs of orthogonally polarized photolithographed planar antennas in each pixel, obviating the need for horns, contacting lenses, or other bulky coupling optics. These planar antenna arrays are composed of slot sub-radiators, spaced to Nyquist sample the focal plane surface to avoid grating lobes. Waves captured by the antenna slots of a given polarization orientation in the pixel are coherently combined through a microstrip summing tree. We control the illumination pattern in each pixel through the microstrip line impedance surrounding each T-junction; in BICEP3, we use a gaussian illumination pattern that minimizes spillover onto the cold aperture stop. Optical power from the antennas passes through an on-chip band pass filter before thermally dissipating on a released bolometer island in close thermal contact to the TESes (Figure ~\ref{fig:TES}). Each bolometer consists of two TESes: an aluminum TES with a transition temperature $T_{c}\sim$\SI{1.2}{K} for higher loading in-lab testing and a titanium TES with $T_{c}\sim$\SI{0.5}{K} for on sky measurement. Ref.\citenum{Ade15} shows design parameter and fabrication procedure of the detectors. Each detector tile has 60 pixels at \SI{95}{G\hertz}, with 4 additional dark pixels at the corners for calibration.

\subsection{Focal Plane Module}
\begin{figure} [ht]
   \begin{center}
   \begin{tabular}{c}
   \includegraphics[height=6cm]{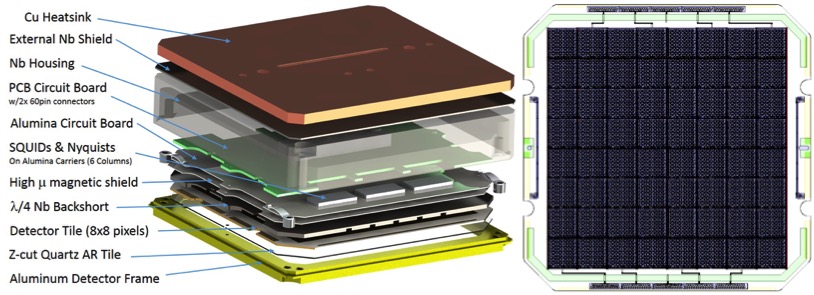}
   \end{tabular}
   \end{center}
   \caption[TES]
   { \label{fig:module_model} 
Left: Solidworks model for the BICEP3 module. The AR tile, detector, backshort and mux circuit board are mounted directly onto the Aluminum frame with tile clips and aligned with pin/slot pairs. Right: mask artwork for detector lithography. Each module contains 60 optically active pixels and 4 dark channels in the corners. All detectors connect to the circuit board via wirebond on top and bottom of the tile.}
   \end{figure}

\begin{figure} [ht]
   \begin{center}
   \begin{tabular}{c}
   \includegraphics[height=7cm]{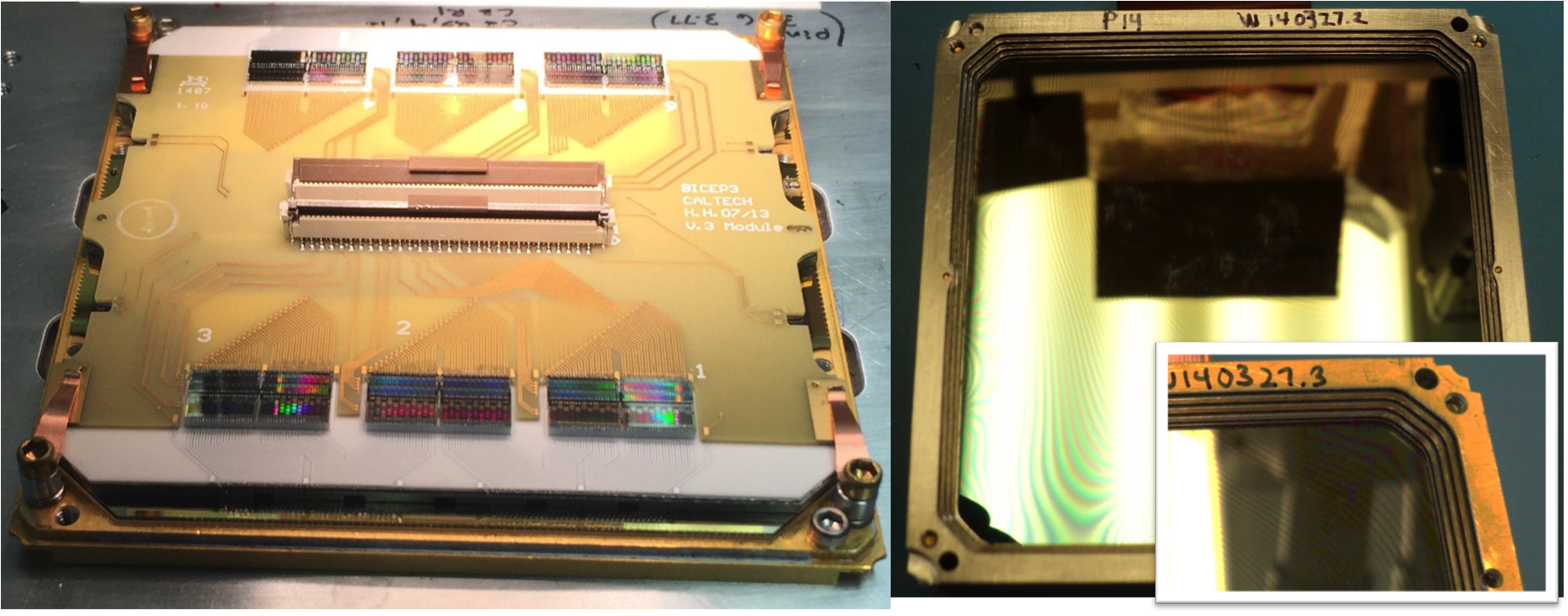}
   \end{tabular}
   \end{center}
   \caption[TES]
   { \label{fig:module_photo} 
Left: Inside of the module. A set of alumina and G-10 circuit boards interconnect the SQUID (MUX) and Nyquist (NYQ) chips with the flex-cable connectors at center. These readout chips connect to the detector tile around a $\lambda/$4 backshort between the detectors and printed circuit boards. A few hundred gold wire bonds directly connect between the detector tile and aluminum detector frame. Right: the front side of the module and the corrugation on the frame to suppress coupling between the edge pixel and the frame. Bottom right: Zoom of the front side to show the corrugation on the frame.}
   \end{figure}

The focal plane module consists of a quartz anti-reflection coating, detector tile, niobium (Nb) quarter-wave backshort, 1st stage superconducting quantum interference device (SQUID) chips, and the readout circuit boards (Figure ~\ref{fig:module_model}). These components are stacked together on an aluminum detector frame, aligned with a \SI{2}{mm} diameter pin/slot pair at opposite side and mounted at the corners with tile clips. The detector tiles thermally sink to their aluminum frames by $\sim$500 gold wire bonds. The front side of the aluminum frame is corrugated to suppress coupling between the edge pixels and the frame (Figure ~\ref{fig:module_photo}).

All the detector parts are enclosed in a niobium housing to control the magnetic field environment near the SQUID ammeters and multiplexers. The module is mounted to a copper heat-sinking piece at the back of the niobium housing only making thermal contact at the center of the niobium housing to avoid trapped magnetic flux during cool down. Every module unit is independently placed onto the \SI{250}{mK} focal plane of the telescope. We mount a low-pass metal mesh filter (developed by Cardiff University \cite{Ade06}) to remove unwanted above-band radiative loading. Two 60 channel flex cables connect to the back of the module.

\begin{figure} [ht]
   \begin{center}
   \begin{tabular}{c}
   \includegraphics[height=5cm]{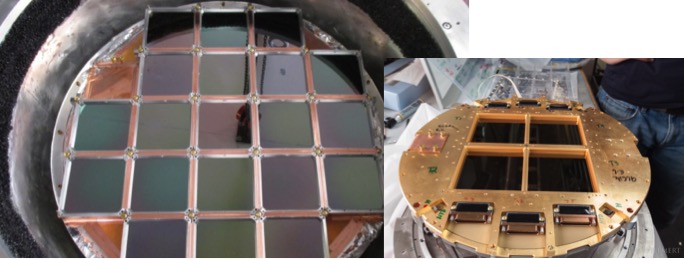}
   \end{tabular}
   \end{center}
   \caption[TES]
   { \label{fig:focal_plane} 
Left: Fully populated BICEP3 focal plane with 20 tiles. Right: Fully populated Keck focal plane with 4 detector tiles. The detector tiles are the same size on both focal plane.}
   \end{figure}

This new modular packaging allows easy repair and upgrades individual detector tiles, and the compact design gives more efficient use of optical area in the focal plane. Together with the faster optics design, BICEP3 has a throughput 10 times higher than that of a single Keck \SI{95}{GHz} receiver (Table ~\ref{tab:B3_throughput}). Figure ~\ref{fig:focal_plane} shows the focal plane currently installed in BICEP3. Future improvement on this design will allow us to minimize its weight and RF/magnetic interference.

   \begin{table}[ht]
\caption{BICEP3 and Keck throughput} 
\label{tab:B3_throughput}
\begin{center}       
\begin{tabular}{|l|l|l|}
\hline
\rule[-1ex]{0pt}{3.5ex}   & Single Keck/B2 & BICEP3 \\
\hline
\rule[-1ex]{0pt}{3.5ex}  Optics & f/2.2 & f/1.68\\
\hline
\rule[-1ex]{0pt}{3.5ex}  Aperture & 264 mm & 520 mm\\
\hline
\rule[-1ex]{0pt}{3.5ex}  FOV & 17 degree  & 27.4 degree \\
\hline
\rule[-1ex]{0pt}{3.5ex}  Throughput & 37.8 cm$^2$sr & 381 cm$^2$sr \\
\hline
\end{tabular}
\end{center}
\end{table}

\subsection{Time-domain Multiplexing and Readout}
BICEP3 uses a time-domain multiplexed (TDM) system developed at NIST for the bolometer readout~\cite{deKorte03}. The readout electronics consists of the Nyquist chips (NYQ), SQUID multiplexing chips (MUX) and the SQUID series array (SSA). The NYQ chips are used to voltage bias the detectors with a \SI{4}{m\ohm} shunt resistor. The chips also have a \SI{2}{$\mu$H} inductors to limit the bandwidth. The MUX chips contain the first stage of the SQUID multiplexer, and the SSAs provide the final SQUID amplifier stage. The NYQ and MUX chips are located inside the module cooled to \SI{280}{mK}, while the SSAs are attached to the \SI{4}{K} temperature stage. A Multi-Channel Electronic (MCE) system developed by the University of British Columbia controls the bias and readout of all the channels~\cite{Battistelli08}.

\begin{figure} [ht]
   \begin{center}
   \begin{tabular}{c}
   \includegraphics[height=8cm]{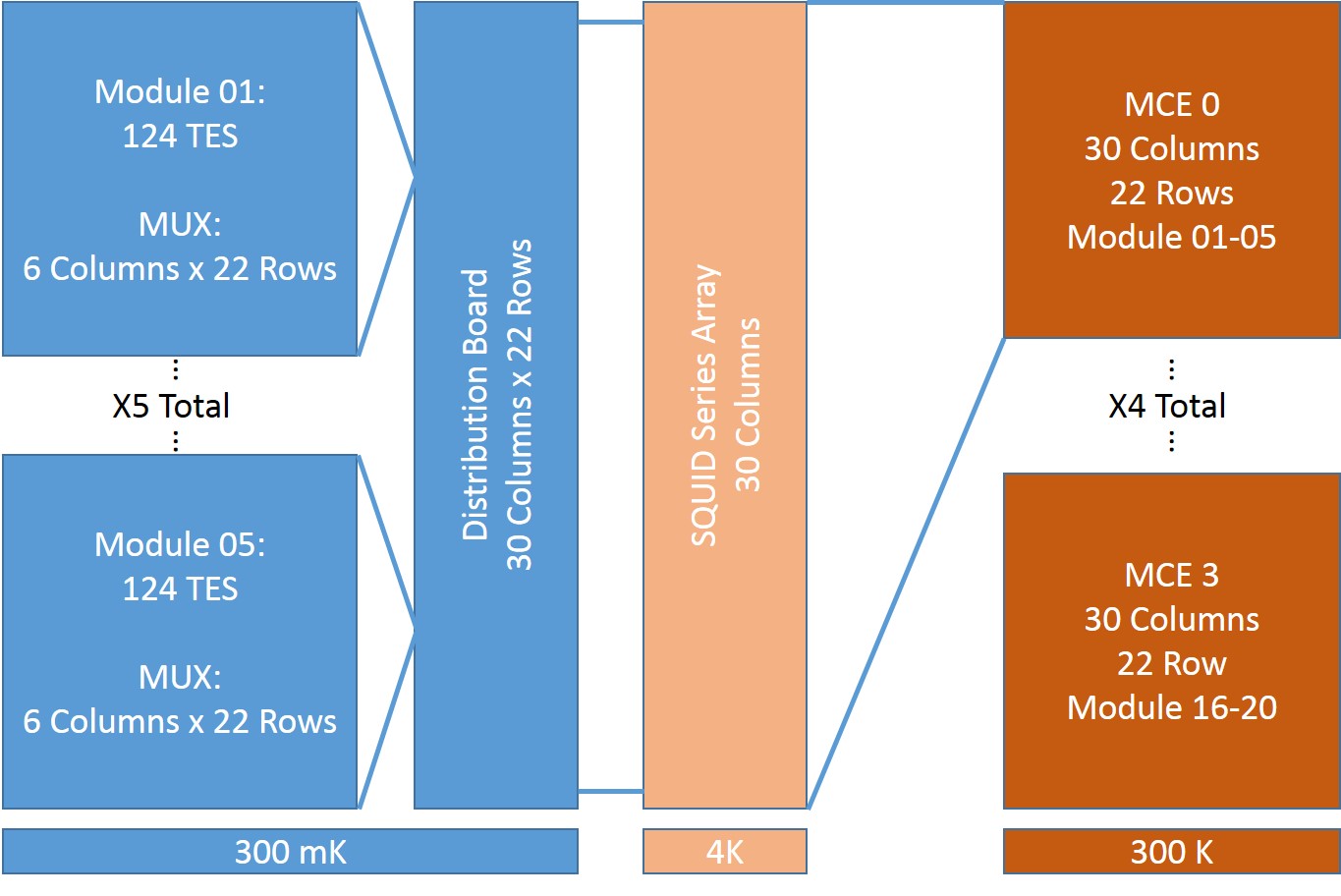}
   \end{tabular}
   \end{center}
   \caption[TES]
   { \label{fig:MCE_block} 
Readout schematic of BICEP3. Every 5 modules are grouped and connected to a distribution board behind the focal plane at \SI{280}{mK}, then to the SSA at \SI{4}{K}, and connected to the room temperature MCE. 4 MCEs are used to readout all 20 tiles (2560 detectors).}
   \end{figure}

The multiplexing architecture is $22\times30\times5$: 22 TESes are read out in a multiplexer row and there are 30 multiplexer columns to form a MCE unit. Each set of the NYQ-MUX chip corresponds to a signal column and 11 rows, 2 chips are connected to form the 22 row multiplexing set, and 6 of these sets are mounted inside each module. 5 modules connect to a circuit board behind the focal plane (distribution board) to group all 30 columns. The row select lines are wired in series for every 5 modules. Superconducting niobium-titanium, twisted-pair cables connect the focal plane and SSAs at \SI{4}{K}. They are readout by a MCE unit attached to the cryostat at \SI{300}{K}. Four independent MCE units read out all 20 modules. Figure ~\ref{fig:MCE_block} shows the block diagram of the readout schematic.

\subsubsection{SQUID Amplifier and Multiplexer}
The SQUIDs play several simultaneous roles in our readout system.  They amplify the small current output of the TESes while adding noise sub-dominant to the TES itself.  They transform the small $\sim$\SI{60}{m\ohm} impedance of the TES to levels that warm amplifiers can match.  Lastly, they have sufficient bandwidth to allow multiplexing of several detectors on common readout lines.

\begin{figure} [ht]
   \begin{center}
   \begin{tabular}{c}
   \includegraphics[height=8.5cm]{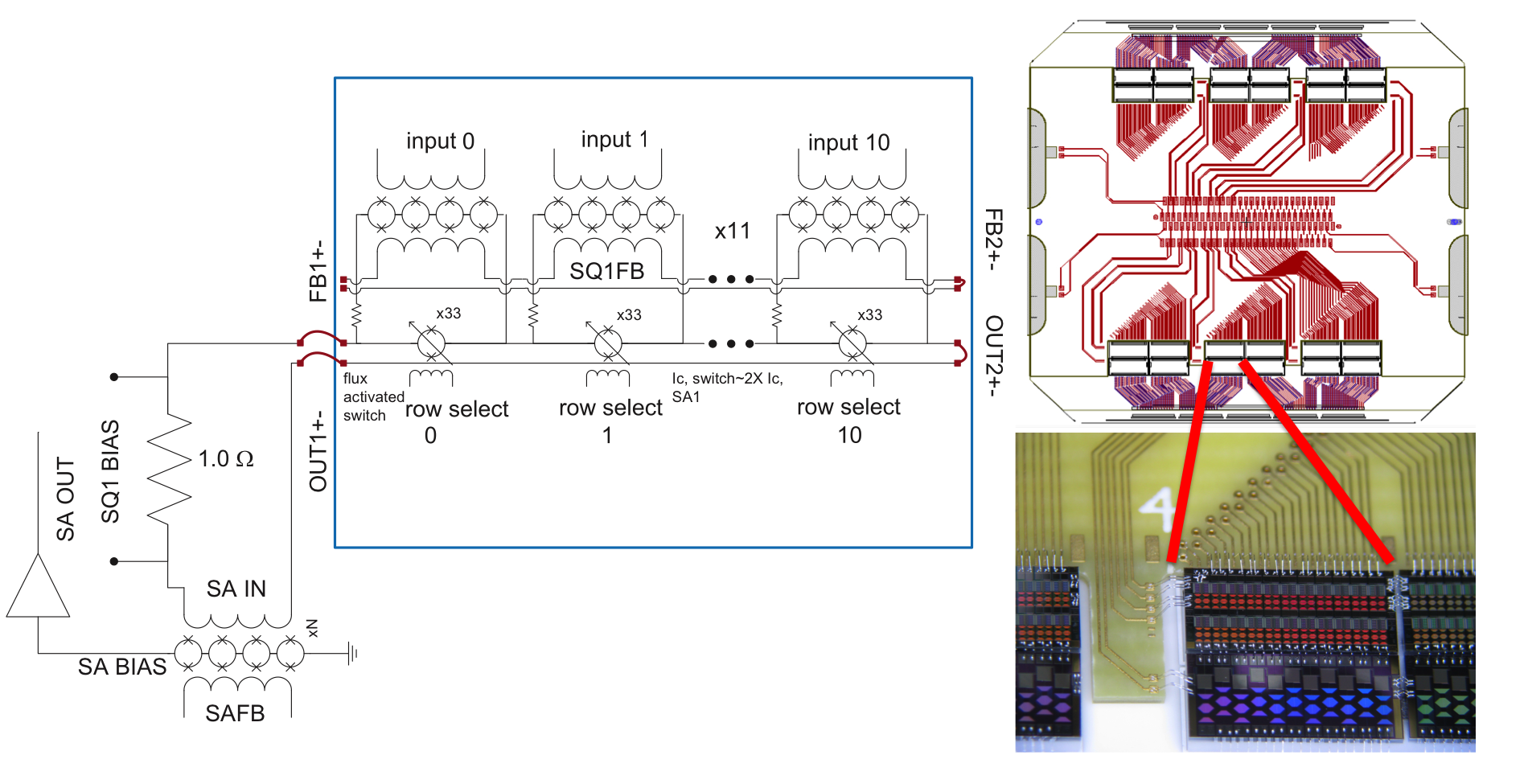}
   \end{tabular}
   \end{center}
   \caption[TES]
   { \label{fig:squid_pic} 
Left: A simplified diagram of of the SQUID MUX used in BICEP3. All elements inside the box are located inside the module, the SA component is sunk to \SI{4}{K}. Right: The circuit layout for the module circuit board and a zoom in of the NYQ/MUX chips. Two 11-row MUX chips are connected together to form a column, and each module has 6 columns.}
   \end{figure}

Each independent detector is inductively coupled to a signal SQUID array (SQ1) by an input coil and the amplifier is operated in flux-lock loop to linearize the periodic output and increase the dynamic range of the SQUIDâs response. As the flux from the input coil changes in response to the TES current, a compensating flux is applied by the feedback coil to cancel it. This flux feedback serves as the output of the TES channel. The SSA provides an additional stage of amplification that provides the aforementioned impedance matching between the first stage SQUIDs and room temperature MCE, providing $\sim$\SI{1}{\ohm} dynamic resistance for a $\sim$\SI{100}{\ohm} output impedance. Figure ~\ref{fig:squid_pic} shows a simplified schematic of the SQUID amplifier system. A similar design is used BICEP2/ Keck and many other experiments.

Time Domain Multiplexing is possible because the SQ1 will not generate output signal when it is biased below its critical current $I_{\text{min}}$. Each SQ1 couples a TES to a shared common readout amplifier (SSA).  While the TESes are continuously biased, they are only sampled when the corresponding SQ1 channel is biased.  This allows our readout system to sequentially read 22 detectors in a common column, revisiting frequently enough to nyquist sample the highest relevant frequencies in the time-stream.

Each SQ1 bias in a signal column is controlled by a superconducting-to-normal flux activated switch that biases in parallel with the SQ1 and is controlled by the 22 row-select (RS) input lines. This design differs from that in BICEP2/Keck where the RS input lines separately biased each row of SQ1s, requiring an extra per-column intermediate summing coil and SQUID (SQ2) before reaching the SSA.

The flux activated switch is designed to switch at twice the critical current of the SQ1s, allowing the switches to share the same bias line with SQ1 in BICEP3. This ultimately reduces the electrical wiring going into the cold stage of the focal plane.

Control of the MUX system and feedback-based readout of the TES data are via the room temperature Multi-Channel Electronics (MCE) systems. The multiplexing speed needs to be quick enough for the Nyquist frequency to exceed the noise bandwidth to avoid aliasing penalty. BICEP2/ Keck shows the optimal multiplexing speed is \SI{25}{k\hertz} with a \SI{2}{$\mu$H} bandwidth limiting inductor~\cite{Kernasovskiy14}. The data are filtered and down sampled in the MCE before being output to the computer software. The MCE uses a fourth-order digital Butterworth filter before down-sampling by a factor of 168, the control software applies a second stage of filtering using an acausal, zero-phase-delay FIR filter to down sampled by another factor of 5, giving a final sample rate of \SI{31.1}{\hertz}. The full multiplexing parameters used in BICEP3 are shown in Table ~\ref{tab:MCE}.

   \begin{table}[ht]
\caption{Summary of multiplexing parameters used by BICEP3.} 
\label{tab:MCE}
\begin{center}       
\begin{tabular}{|l|l|}
\hline
\rule[-1ex]{0pt}{3.5ex}  Raw ADC sample rate & \SI{50}{M\hertz} \\
\hline
\rule[-1ex]{0pt}{3.5ex}  Row dwell & 90 samples \\
\hline
\rule[-1ex]{0pt}{3.5ex}  Row switching rate & \SI{556}{k\hertz} \\
\hline
\rule[-1ex]{0pt}{3.5ex}  Number of rows & 22 \\
\hline
\rule[-1ex]{0pt}{3.5ex}  Sample-row revisit rate& \SI{25.3}{k\hertz} \\
\hline
\rule[-1ex]{0pt}{3.5ex}  Internal downsample& 168 \\
\hline
\rule[-1ex]{0pt}{3.5ex}  Output data rate per channel& \SI{150}{\hertz} \\
\hline
\rule[-1ex]{0pt}{3.5ex}  Software downsample& 5 \\
\hline
\rule[-1ex]{0pt}{3.5ex}  Archived data rate& \SI{31.1}{\hertz} \\
\hline
\end{tabular}
\end{center}
\end{table}

\section{DETECTOR PERFORMANCE}
BICEP3 was first deployed to South Pole during the 2014-15 austral summer with 9 out of 20 detector modules. This season many major improvements were made including sub-Kelvin fridge hold time, RF shielding, IR thermal filtering \cite{Grayson16}, and fully populating the focal plane. 

\subsection{Detector yield}
While the fully populated focal plane had 1200 optically active dual-polarized detector pairs, the final working count is 951 pairs. This is largely due to electrical opens in 7 of the multiplexing rows, likely caused by wire bond damage during cooling.

\subsection{Beams and detector spectral response}
Far-field detector response was measured using a chopped thermal source about \SI{200}{m} from the telescope. Differential pointing, beam width and ellipticity were measured for each detector to characterize and control for beam systematics in analysis. Ref.~\citenum{Karkare16} shows the experimental setup and beam measurement result.

\begin{figure} [ht]
   \begin{center}
   \begin{tabular}{c}
   \includegraphics[height=6cm]{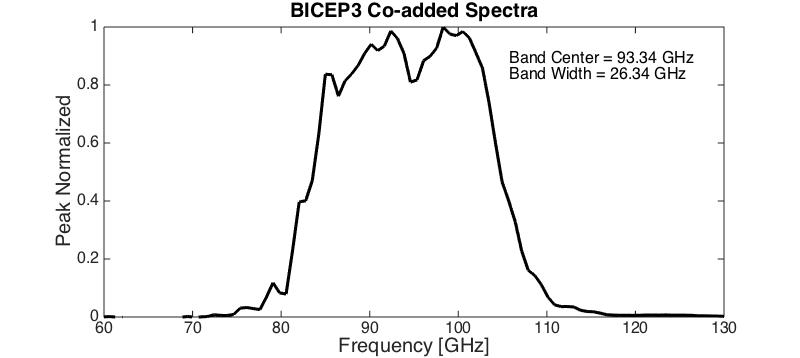}
   \end{tabular}
   \end{center}
   \caption[TES]
   { \label{fig:bicep3_fts} 
Co-added BICEP3 FTS measurement.}
   \end{figure}

The detector spectral response of BICEP3 was measured with a Martin-Puplett interferometer~\cite{Karkare14}. BICEP3 is designed to have an optical band centered at \SI{93}{G\hertz} with a fractional bandwidth of 27\% to avoid the oxygen line at \SI{118}{G\hertz} and below \SI{63}{G\hertz}. Figure ~\ref{fig:bicep3_fts} shows the co-added measured spectra for BICEP3 with a median band center at \SI{93.3}{G\hertz} and bandwidth of \SI{26.3}{G\hertz} (28\% fractional bandwidth). There are evidences of tile to tile non-uniformity spectral respond in this season's measurement that require further investigation.

\subsection{Optical Efficiency}
\begin{figure} [ht]
   \begin{center}
   \begin{tabular}{c}
   \includegraphics[height=6cm]{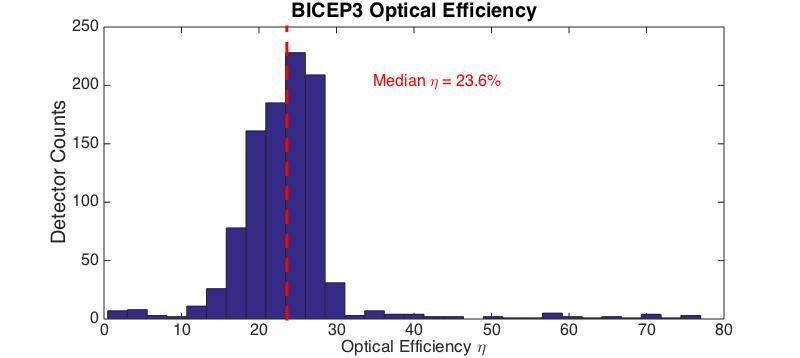}
   \end{tabular}
   \end{center}
   \caption[TES]
   { \label{fig:bicep3_oe} 
Histogram of the detector optical efficiency.}
   \end{figure}

A flat Eccosorb sheet was placed over the cryostat window to act as a beam-filling Rayleigh-Jeans source. It can be cooled with liquid nitrogen to \SI{77}{K} or left at room temperature and we use the detector load curves taken at different temperatures to determine end-to-end optical efficiency (figure ~\ref{fig:bicep3_oe}). The median optical efficiency is $\SI{0.08}{pW/K_{rj}}$, which corresponds to an efficiency $\eta \sim 24 \% $.

\subsection{CMB Map}
\begin{figure} [ht]
   \begin{center}
   \begin{tabular}{c}
   \includegraphics[height=15cm]{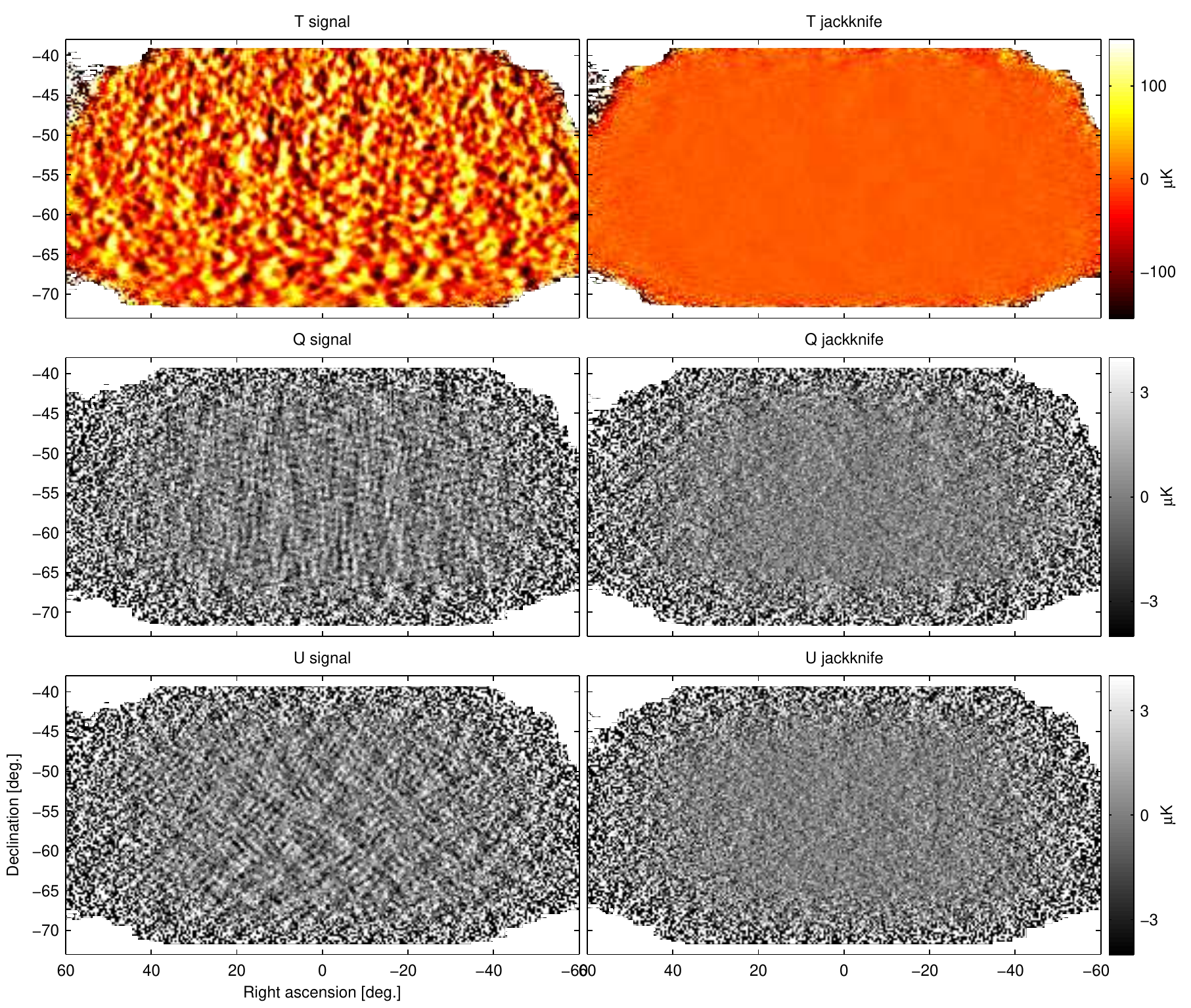}
   \end{tabular}
   \end{center}
   \caption[TES]
   { \label{fig:bicep3_map} 
Preliminary temperature and polarization maps of the CMB made with small amount of obtained data, corresponding to 700 hours of integration time.}
   \end{figure}

We show preliminary CMB temperature and polarization maps using the first 700 hours of 2016 BICEP3 CMB data (figure ~\ref{fig:bicep3_map}) with detector pointing and absolute gain calibrated by correlation with a reference Planck temperature map.

\subsection{Preliminary Map-based NET}
\begin{figure} [ht]
   \begin{center}
   \begin{tabular}{c}
   \includegraphics[height=7cm]{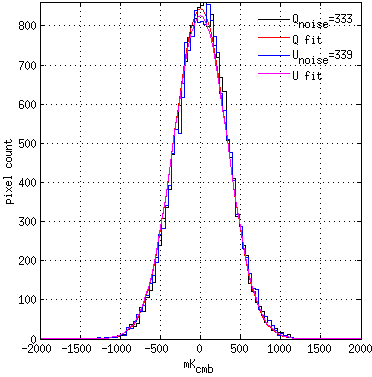}
   \end{tabular}
   \end{center}
   \caption[TES]
   { \label{fig:bicep3_net} 
Per-detector map-based NET of BICEP3 derived using the first 700 hours of CMB data in 2016 season.}
   \end{figure}

Noise of the telescope can be estimate by evaluating the polarization map in section 3.4. The difference of two polarization maps made with data taken in the opposite azimuth scan directions (forward/backward jackknife) was used to remove CMB signal. The per-detector NET was calculated by multiplying the noise of the jackknife polarization maps and the square root of the integration time map. Figure ~\ref{fig:bicep3_net} shows the histogram of the time weighted noise, which is well described by a gaussian distribution. 

The median per-detector NET is \SI{333}{\ukrts} (figure ~\ref{fig:bicep3_net}) and the telescope NET is \SI{9.72}{\ukrts}. This is a major improvement from the first season, which had a median NET of \SI{395}{\ukrts}. This is still 28\% higher than Keck \SI{95}{G\hertz} receivers at about \SI{260}{\ukrts}, possibly due to excess optical loading from the telescope. 

This sensitivity estimation is different than the one used in Ref.~\citenum{Grayson16}, which calculated the NET by evaluating the time-stream data. Data were first pair-differenced, subjected to a third-order polynomial filter to reduce the $1/f$ noise induced from the atmosphere, and condensed into a noise spectra. The noise performance of the detector is then evaluated by taking the median of the noise spectrum in the science band from 0.1 - \SI{1}{\hertz}. The time-stream based calculation gives per-detector NET at \SI{347}{\ukrts} and receiver NET at \SI{9.91}{\ukrts}. The map-based NET calculation is lower than the time-stream method due to smaller weighing in the low frequency noise.

\section{CONCLUSION}
In this proceedings, we present the the design of the BICEP3 focal plane module and its readout architecture. This compact design increases the packing density of the detectors and allows more efficient use of optically illuminated area on the focal plane. The modular design makes future replacements and upgrades easier. We also show great improvement in detector performance in the second season of BICEP3, increasing detector yield from 436 to 951 
polarization-sensitive pixels, reducing the per-detector NET from \SI{395}{\ukrts} to \SI{333}{\ukrts}, and achieving a receiver NET of \SI{9.72}{\ukrts}.

\section{ACKNOWLEDGMENTS}
The Bicep3 project has been made possible through support from the National Science Foundation (grant Nos. 0742818, 0742592, 1044978, 1110087, 1145172, 1313158, 1313010, 1313062, 1313287, 1056465, and 0960243), the W. M. Keck Foundation, the Canada Foundation for Innovation, and the British Columbia Development Fund. The development of antenna-coupled detector technology was supported by the JPL Research and Technology Development Fund and grants 06-ARPA206-0040 and 10-SAT10-0017 from the NASA ARPA and SAT programs. The development and testing of focal planes were supported by the Gordon and Betty Moore Foundation at Caltech. The computations in these proceedings were run on the Odyssey cluster supported by the FAS Science Division Research Computing Group at Harvard University. Tireless administrative support was provided by Irene Coyle, Kathy Deniston, Donna Hernandez, and Dana Volponi. 

We are grateful to the staff of the US Antarctic Program and in particular the South Pole Station without whose help this research would not have been possible. We thank our Bicep1, Bicep2, Keck Array and Spider colleagues for useful discussions and shared expertise.
\bibliography{report} 

\begin{thebibliography}{10}

\bibitem{Kamionkowski15}
{Kamionkowski}, M. and {Kovetz}, E.~D., ``{The Quest for B Modes from
  Inflationary Gravitational Waves},'' {\em ArXiv}  (Oct. 2015).

\bibitem{Ogburn12}
{Ogburn}, R.~W., {Ade}, P.~A.~R., {Aikin}, R.~W., {Amiri}, M., {Benton}, S.~J.,
  {Bischoff}, C.~A., {Bock}, J.~J., {Bonetti}, J.~A., {Brevik}, J.~A.,
  {Bullock}, E., {Burger}, B., {Davis}, G., {Dowell}, C.~D., {Duband}, L.,
  {Filippini}, J.~P., {Fliescher}, S., {Golwala}, S.~R., {Gordon}, M.,
  {Halpern}, M., {Hasselfield}, M., {Hilton}, G., {Hristov}, V.~V., {Hui}, H.,
  {Irwin}, K., {Kaufman}, J.~P., {Keating}, B.~G., {Kernasovskiy}, S.~A.,
  {Kovac}, J.~M., {Kuo}, C.~L., {Leitch}, E.~M., {Lueker}, M., {Montroy}, T.,
  {Netterfield}, C.~B., {Nguyen}, H.~T., {O'Brient}, R., {Orlando}, A.,
  {Pryke}, C.~L., {Reintsema}, C., {Richter}, S., {Ruhl}, J.~E., {Runyan},
  M.~C., {Schwarz}, R., {Sheehy}, C.~D., {Staniszewski}, Z.~K., {Sudiwala},
  R.~V., {Teply}, G.~P., {Thompson}, K., {Tolan}, J.~E., {Turner}, A.~D.,
  {Vieregg}, A.~G., {Wiebe}, D.~V., {Wilson}, P., and {Wong}, C.~L., ``{BICEP2
  and Keck array operational overview and status of observations},'' in [{\em
  Millimeter, Submillimeter, and Far-Infrared Detectors and Instrumentation for
  Astronomy VI}{\nolinebreak\hspace{0.1em}]},  {\em Proc. SPIE} {\bf 8452},
  84521A (Sept. 2012).

\bibitem{Wu15}
Wu, W. L.~K., Ade, P. A.~R., Ahmed, Z., Alexander, K.~D., Amiri, M.,
  D.~Barkats, S. J.~B., Bischoff, C.~A., Bock, J.~J., Bowens-Rubin, R., Buder,
  I., Bullock, E., Buza, V., Connors, J.~A., Filippini, J.~P., Fliescher, S.,
  Grayson, J.~A., Halpern, M., Harrison, S., Hilton, G.~C., Hristov, V.~V.,
  Hui, H., Irwin, K.~D., Kang, J., Karkare, K.~S., Karpel, E., Kefeli, S.,
  Kernasovskiy, S.~A., Kovac, J.~M., Kuo, C.~L., Megerian, K.~G., Netterfield,
  C.~B., Nguyen, H.~T., O’Brient, R., Ogburn, R.~W., Pryke, C., Reintsema,
  C.~D., Richter, S., Sorensen, C., Staniszewski, Z.~K., Steinbach, B.,
  Sudiwala, R.~V., Teply, G.~P., Thompson, K.~L., Tolan, J.~E., Tucker, C.~E.,
  Turner, A.~D., Vieregg, A.~G., Weber, A.~C., Wiebe, D.~V., Willmert, J., and
  Yoon, K.~W., ``Initial performance of {BICEP}3: A degree angular scale 95
  {GH}z band polarimeter,'' {\em Journal of Low Temperature Physics}  (2015).

\bibitem{Grayson16}
Grayson, J.~A., ``{BICEP}3 performance overview and future prospects for a
  multi-receiver array,'' in [{\em These
  proceedings}{\nolinebreak\hspace{0.1em}]},  {\em Proc. SPIE} {\bf 9914}
  (2016).

\bibitem{Ade15}
Ade, P. A.~R., Aikin, R.~W., Amiri, M., Barkats, D., Benton, S.~J., Bischoff,
  C.~A., Bock, J.~J., Bonetti, J.~A., Brevik, J.~A., Buder, I., Bullock, E.,
  Chattopadhyay, G., Davis, G., Day, P.~K., Dowell, C.~D., Duband, L.,
  Filippini, J.~P., Fliescher, S., Golwala, S.~R., Halpern, M., Hasselfield,
  M., Hildebrandt, S.~R., Hilton, G.~C., Hristov, V., Hui, H., Irwin, K.~D.,
  Jones, W.~C., Karkare, K.~S., Kaufman, J.~P., Keating, B.~G., Kefeli, S.,
  Kernasovskiy, S.~A., Kovac, J.~M., Kuo, C.~L., LeDuc, H.~G., Leitch, E.~M.,
  Llombart, N., Lueker, M., Mason, P., Megerian, K., Moncelsi, L., Netterfield,
  C.~B., Nguyen, H.~T., OâBrient, R., IV, R. W.~O., Orlando, A., Pryke,
  C., Rahlin, A.~S., Reintsema, C.~D., Richter, S., Runyan, M.~C., Schwarz, R.,
  Sheehy, C.~D., Staniszewski, Z.~K., Sudiwala, R.~V., Teply, G.~P., Tolan,
  J.~E., Trangsrud, A., Tucker, R.~S., Turner, A.~D., Vieregg, A.~G., Weber,
  A., Wiebe, D.~V., Wilson, P., Wong, C.~L., Yoon, K.~W., Zmuidzinas, J., and
  {for the Bicep2 , Keck Array, and Spider Collaborations}, ``Antenna-coupled
  tes bolometers used in bicep2, keck array, and spider,'' {\em The
  Astrophysical Journal}~{\bf 812}(2),  176 (2015).

\bibitem{Ade06}
Ade, P. A.~R., Pisano, G., Tucker, C., and Weaver, S., ``{A review of metal
  mesh filters},'' in [{\em Society of Photo-Optical Instrumentation Engineers
  (SPIE) Conference Series}{\nolinebreak\hspace{0.1em}]},  {\em Proc. SPIE}
  {\bf 6275},  62750U (June 2006).

\bibitem{deKorte03}
de~Korte, P. A.~J., Beyer, J., Deiker, S., Hilton, G.~C., Irwin, K.~D.,
  MacIntosh, M., Nam, S.~W., Reintsema, C.~D., Vale, L.~R., and Huber, M.~E.,
  ``Time-division superconducting quantum interference device multiplexer for
  transition-edge sensors,'' {\em Review of Scientific Instruments}~{\bf
  74}(8),  3807--3815 (2003).

\bibitem{Battistelli08}
{Battistelli}, E.~S., {Amiri}, M., {Burger}, B., {Halpern}, M., {Knotek}, S.,
  {Ellis}, M., {Gao}, X., {Kelly}, D., {Macintosh}, M., {Irwin}, K., and
  {Reintsema}, C., ``{Functional Description of Read-out Electronics for
  Time-Domain Multiplexed Bolometers for Millimeter and Sub-millimeter
  Astronomy},'' {\em Journal of Low Temperature Physics}~{\bf 151},  908--914
  (May 2008).

\bibitem{Kernasovskiy14}
Kernasovskiy, S., {\em Measuring the polarization of the cosmic microwave
  background with the Keck Array and Bicep2}, PhD thesis, Stanford University
  (10 2014).

\bibitem{Karkare16}
Karkare, K.~S., ``Optical characterization of the {BICEP}3 {CMB} polarimeter at
  the south pole,'' in [{\em These proceedings}{\nolinebreak\hspace{0.1em}]},
  {\em Proc. SPIE} {\bf 9914} (2016).

\bibitem{Karkare14}
Karkare, K.~S., Ade, P. A.~R., Ahmed, Z., Aikin, R.~W., Alexander, K.~D.,
  Amiri, M., D.~Barkats, S. J.~B., Bischoff, C.~A., Bock, J.~J., Bonetti,
  J.~A., Brevik, J.~A., Buder, I., Bullock, E., Burger, B., Connors, J.~A.,
  Crill, B.~P., Davis, G., Dowell, C.~D., Duband, L., Filippini, J.~P.,
  Fliescher, S., Golwala, S.~R., Gordon, M.~S., Grayson, J.~A., Halpern, M.,
  Hasselfield, M., Hildebrandt, S.~R., Hilton, G.~C., Hristov, V.~V., Hui, H.,
  Irwin, K.~D., Kang, J., Karpel, E., Kefeli, S., Kernasovskiy, S.~A., Kovac,
  J.~M., Kuo, C.~L., Leitch, E.~M., Lueker, M., Mason, P., Megerian, K.~G.,
  Netterfield, C.~B., Nguyen, H.~T., O’Brient, R., Ogburn, R.~W., Pryke, C.,
  Reintsema, C.~D., Richter, S., Schwarz, R., Sheehy, C.~D., Staniszewski,
  Z.~K., Sudiwala, R.~V., Teply, G.~P., Thompson, K.~L., Tolan, J.~E., Turner,
  A.~D., Vieregg, A.~G., Weber, A.~C., Wong, C.~L., Wu, W. L.~K., and Yoon,
  K.~W., ``Keck array and {BICEP}3: spectral characterization of 5000+
  detectors,'' in [{\em Millimeter, Submillimeter, and Far-Infrared Detectors
  and Instrumentation for Astronomy VII}{\nolinebreak\hspace{0.1em}]},  {\em
  Society of Photo-Optical Instrumentation Engineers (SPIE) Conference Series}
  {\bf 9153} (2014).

\end{thebibliography}
\bibliographystyle{spiebib} 

\end{document}